\documentclass{article}

\PassOptionsToPackage{numbers, compress}{natbib}

\usepackage[preprint]{neurips_2022}

\usepackage[utf8]{inputenc} 
\usepackage[T1]{fontenc}    
\usepackage{hyperref}
\hypersetup{colorlinks,allcolors=black}
\usepackage{url}            
\usepackage{booktabs}       
\usepackage{amsfonts}       
\usepackage{nicefrac}       
\usepackage{microtype}      
\usepackage{xcolor}         
\usepackage{graphicx}
\usepackage{wrapfig}
\usepackage{float}
\usepackage{amsmath}
\usepackage{multicol}
\usepackage{multirow}
\usepackage{subfigure}
\usepackage{caption}

\title{ fMRI from EEG is only Deep Learning away: the use of interpretable DL to unravel EEG-fMRI relationships
}

\author{%
  Alexander Kovalev\\
Center for Bioelectric Interfaces, \\ Higher School of Economics, Moscow, Russia \\  \texttt{koval.alvi@gmail.com} \\
   \And
   Ilia Mikheev\\
    Center for Bioelectric Interfaces, \\ Higher School of Economics, Moscow, Russia \\   \texttt{ilia.mikheev@hse.ru} \\
   \And
   Alexei Ossadtchi  \\
Center for Bioelectric Interfaces, Higher School of Economics, \\ 
Artificial Intelligence Research Institute, AIRI, Moscow, Russia  \\ \texttt{ossadtchi@gmail.com} \\
}

\begin{document}

\maketitle

\begin{abstract}

The access to activity of subcortical structures offers unique opportunity for building intention dependent brain-computer interfaces, renders abundant options for exploring a broad range of cognitive phenomena in the realm of affective neuroscience including complex decision making processes and the eternal free-will dilemma and facilitates diagnostics of a range of neurological deceases. So far this was possible only using bulky, expensive and immobile fMRI equipment. 
Here we present an interpretable domain grounded solution to recover the activity of several subcortical regions from the multichannel EEG data and demonstrate up to 60 $\%$ correlation between the actual \textit{subcortical} blood oxygenation level dependent (\textit{s}BOLD) signal and its EEG-derived twin. Then, using the novel and theoretically justified weight interpretation methodology we recover individual spatial and time-frequency patterns of scalp EEG predictive of the hemodynamic signal in the subcortical nuclei. 
The described results not only pave the road towards wearable subcortical activity scanners but also showcase an automatic knowledge discovery process facilitated by deep learning technology in combination with an interpretable domain constrained architecture and the appropriate downstream task. 

\end{abstract}

\section{Introduction}
Electroencephalography(EEG) and functional magnetic resonance imaging (fMRI) are unarguably the two most frequently used modalities to non-invasively register the brain activity. These techniques are grounded into very different principles and have complementary spatial and temporal resolution properties.  

This  complementary nature of the aforementioned brain imaging modalities and their resolution properties lead to numerous attempts of developing methods for solving EEG (and MEG) inverse problems based on the \textit{apriori} information extracted from the fMRI.  Within this approach the activity maps discovered by fMRI are used to form probabilistic priors to regularize and reduce the originally ill-posed EEG and MEG inverse problem \cite{megfmri}. Also, EEG triggered fMRI appeared useful for localization of seizure onset zone in patients with epilepsy, \cite{gotman2011combining}.  

Another promising way to combine the two imaging modalities is to use the Representational Similarity Analysis \cite{rsa} to benefit from high temporal resolution of electrophysiological mapping and superior spatial resolving properties of the BOLD based technique. The approach described in \cite{cichy} is based on the brain activation signals induced by the same of stimuli and separately recorded by the two modalities. This powerful technique completely bypasses the questions related to mutual relationship between the biophysical and physiological aspects of the two imaging techniques and relies exclusively on the comparing the differences in reaction of the brain to a set of stimuli, called representation patterns, as meausured by either EEG/MEG or fMRI.  The results of this technique crucially depend on the set of stimuli used to evoke brain responses and therefore, this method can not be applied to combine the BOLD and electric signals in the non-stimulus based paradigms, such as those exploring resting state brain activity \cite{restingstate}.

At the same time, the fMRI is famous for the studies exploring dynamics of resting state networks (RSN) and discovery of several default mode networks (DMN) comprising specific hubs that mediate the informational exchange between brain structures \cite{rsnfmri}.  However, due to the low temporal resolution of fMRI  most of the observations reflect mere similarity of slowly varying activation patterns of distal neuronal populations, so that the fine temporal structure accompanying such synchronous activity appears lost.  

Most of the known to date correlates of cognitive processes were discovered using fMRI. It has always been tempting to use this information obtained by means of a very expensive and immobile equipment and create affordable and compact solutions aimed at monitoring activity of the specific brain structures highlighted in the fMRI. The most readily available option is to build an EEG based solutions based on monitoring of the electric activity of the neural populations located in the fMRI discovered hotspots. As it was done in \cite{medi}, only spatial information is used and the spatial filter is created to be applied to the multichannel EEG data in order to estimate the electrical activity of the fMRI discovered cortical region.

This approach absolutely ignores the intricate relationships between the electric and  BOLD signals. For example in \cite{hemoeeg} a heuristic is suggested that links BOLD signal to the increase of the root mean square frequency in the EEG signal. Also, it was shown that BOLD is correlated with gamma EEG activity and also with non-linear phase-amplitude coupling between theta and gamma bands. Also, when such fMRI informed EEG solvers are used in a real-time setting it is important to take into account the timing differences between the EEG and fMRI. It is a general belief that  FMRI signal is not only blurred  in time  but is also delayed by some 4-8 s with respect to the EEG measurements.   

A relatively recent  revolutionary paradigm allows for simultaneous recording of EEG and fMRI data \cite{eegfmrirev}, \cite{mbfmri}. This possibility triggered a range of studies aimed at relating fMRI and EEG signals in various experimental paradigms. The majority of attempts establish the similarity of EEG and fMRI based findings about the underlying brain activity by  directly correlating the two measures with prior trasformation of the EEG data to the fMRI scale by a simple convolution of EEG with Hemodynamic Response Function , e.g. \cite{eegfmrialpha}, \cite{concfmrieeg}.  
 
A more advanced approach has been describde in \cite{sato2010} where the authors developed a method based on finding the voxel-by-voxel EEG to fMRI mapping functions. They demonstrate  that it is possible to obtain meaningful localization by using BOLD signal estimated  from the simultaneously recorded  fMRI and EEG data. There was an attempt to predict the components of single trial EEG response from fMRI data. In \cite{martino2011} the authors used Relevance Vector Machine Regression  to build a predictor that allowed to solve the inherently ill-posed task of mapping EEG from fMRI signal in a stimulus based paradigm. Leite in \cite{eegfmriepi} attempted to build a transfer function between EEG and BOLD signals in the context of analysis of epileptic activity. They made an important observation that frequency weighted data yield better prediction accuracy than that obtained without taking into account the band-wise distribution of electric activity. The focus of the previous work was on building the decoders that can predict the data in one modality from the measurement obtained within the other.

Dominantly, the studies exploring EEG-BOLD relationship focus on the analysis if cortical activity and not the subcortical structures such as for example basal ganglia, parts of the brain responsible for a broad range of behaviours, reflecting emotional state, underlying complex decision making and risk-reward balancing processes, voluntary movements control and various kinds of learning. 

It has been recently shown that cortical BOLD signal can be reconstructed from multichannel EEG data by means of machine learning \cite{ml_eeg_fmri}. Although fascinating and insightful regarding BOLD-EEG relationship, from the practical standpoint this result is of limited value since cortical activity can be estimated from EEG using classical inverse modelling machinery \cite{fmri_prior} without the need to pretrain the decoder. 

Despite few reports \cite{eeg_informed, review_eeg_fmri}, recovery of subcortical BOLD (sBOLD) signal from EEG remains elusive.  Building such a decoder with interpretable and domain grounded neural networks would allow for additional insights regarding the connection between the subcortical BOLD and cortex-wide electrical activity and will the road towards creation of wearable and low-cost scanners of subcortical activity with a broad range of applications from fundamental neuroscience research to neurorehabilitation. Here we attempted to build such a decoder that would predict the BOLD signal of several subcortical structures using the multichannel EEG data. Based on the previous work \cite{ml_eeg_fmri_2, ml_eeg_fmri}, we hypothesized that the sBOLD signal can be reconstructed using instantaneous band power values in the corresponding spatially filtered EEG signal. Our decoder's front end comprises well accepted in the EEG data processing field spatial and frequency domain selective filters followed by the non-linearity and the smoother to extract the instantaneous power of the underlying neuronal sources. Each such spatial filter in fact solves the inverse problem for a specific brain region or their combination and corresponds to the rows of an inverse operator. However, unlike in the classical approach the coefficients of such a filter are found in an adaptive fashion in the context of the given downstream task of matching the actual BOLD and its EEG-derived twin. The filters formed through training not only get tuned to the target activity but also tune way from the interfering signals \cite{haufe2014interpretation}, something which needs to be taken into account when interpreting the weights. Being serially cascaded with a temporal filter the spatial filter adapts and tunes away in the context set by the corresponding temporal filter and vice versa  \cite{petrosyan2021decoding}. Weights interpretation then allows us to assess the extent to which the obtained solution is physiological and extract the new knowledge about the sBOLD-EEG relationship.  

\section{Method}
\subsection{CWL Dataset}

We perform experiments on the “Eyes Open – Eyes Closed” EEG/fMRI dataset with Carbon Wire Loop channels(CWL dataset) \cite{cwl_data}. In this dataset, electroencephalography (EEG) and functional magnetic resonance imaging fMRI measured BOLD signals are simultaneously recorded. The dataset comprises records of the resting state data in eight volunteers for five minutes each. The first four subjects were recorded by Trio MRI scanner with a time of repetition (TR) of 1.95 seconds. The other four subjects were scanned by Verio scanner, TR = 2.00 seconds. A carbon wire loop (CWL) was added to the EEG cap during the EEG/fMRI experiment. Signals induced by the gradient coils in the CWL were adaptively subtracted from the EEG leads to reduce the amount of gradient artifacts in the measurements of brain's electrical activity. The EEG was recorded using a 30 channel MR-compatible electrode cap with a native sampling frequency of 5000 Hz during the fMRI scan. Signals were corrected for gradient artifacts and then downsampled to sampling frequency 1000 Hz by the authors of this dataset. BOLD signals consist of 150 three-dimensional tensors with shape $61\times72\times61$ with resolution 3×3×3 mm. We then normalized the BOLD data to the MNI space and slice-time corrected.

\subsection{Preprocessing}
\begin{figure}[ht]
    \centering
    \includegraphics[width=0.9\textwidth]{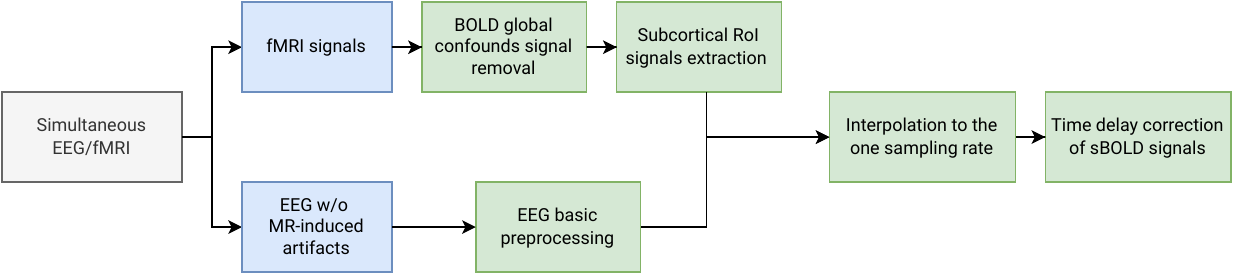}
    \caption{\label{fig:preproc} \textbf{EEG-fMRI data preprocessing.}. fMRI signals and multichannel EEG without gradient artifacts are taken from the CWL dataset. Next, relevant brain activity is extracted from subcortical regions of BOLD signals. Basic preprocessing and normalization is applied to EEG data. Finally, interpolation and time-delayed corrections to sBOLD signals are applied.}
\end{figure}

We preprocessed the data according  to the diagram presented in Figure \ref{fig:preproc}.

\paragraph{EEG processing}
The EEG data cleaned from the gradient and ballistocardiographic artifacts were notch filtered to remove power-line signal and its harmonics at 50, 100 and 150Hz. Then we have also applied a broadband filter with 1-100 Hz passband to remove the extraneous noise and lower frequency drifts. We used FIR filter with window size 6.601 s. for notch filter and 3.301 s for  for broadband filter. Then we re-referenced the multichannel EEG data to the common average reference montage. 

\paragraph{fMRI processing} 
In order to reduce noise in BOLD signals, a spatial Gaussian filter  with Full-Width at Half Maximum(FWHM) = 3 mm was applied. Then we removed signals that did not relate to neural activity. To this end we applied confound regression using motion signals and global signals as regressors, and subtracted those signals from BOLD activity. This reduced the interdependence between activity of different regions. Subsequently, we extracted eight subcortical regions from voxels to reduce the dimensionality of the BOLD data using Harvard-Oxford subcortical structural atlas \cite{oxford_atlas}.  We decided to focus on the following four bilaterally symmetric basal ganglia regions: pallidum, caudate, putamen and accumbens.

Finally, the preprocessed EEG were downsampled and the subcrotical ROI BOLD (sBOLD) signals were interpolated to the common sampling rate of 100 Hz. While EEG directly reflects manifestations of the electrical activity of large neuronal populations, the sBOLD data represent hemodynamic response and have very different properties as compared to the EEG. To relate BOLD to EEG a hemodynamic response function (HRF) \cite{hrf_bold} is typically used and applied to the smoothed power profiles of EEG derived cortical activity. The HRF has a characteristic peak in the 4-10 seconds range which reflects the fact the hemodynamic activity is typically delayed with respect to the electrical signal. To account for this we corrected our data for BOLD time delay. Therefore, each EEG time point $t_{eeg}$ corresponds to each BOLD time point $t_{bold} = t_{eeg} + t_{delay}$, where $t_{delay}$ is the hyperparameter which we tune.

\subsection{The model architecture}

\begin{figure}[ht]
    \centering
    \includegraphics[width=1\textwidth]{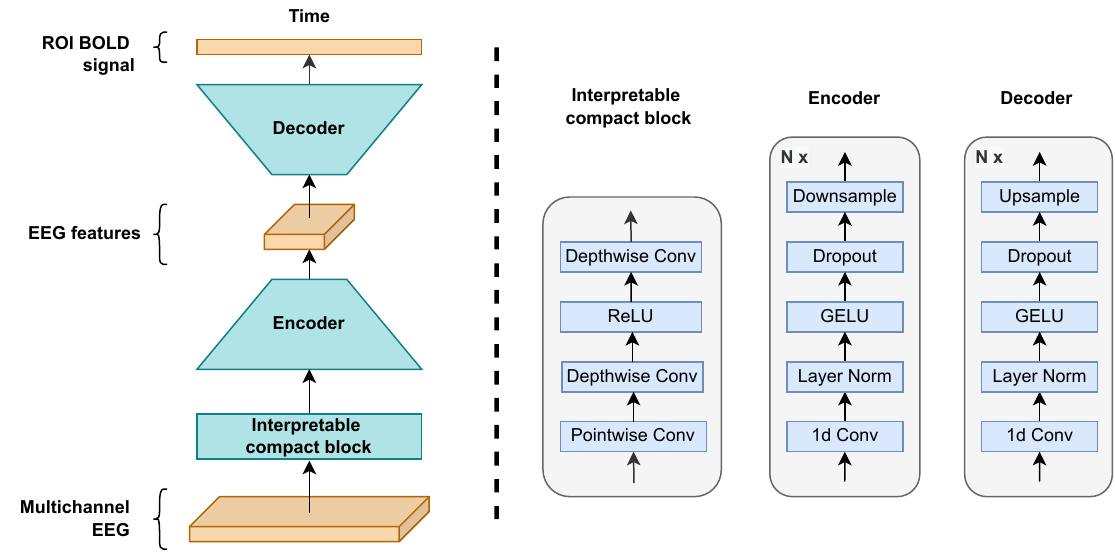}
    \caption{\label{fig:model}(Left) \textbf{BOLD from EEG Interpretable Regression Autoencoder.} 
    We apply our architecture to a block of raw EEG data. First, the multibranch interpretable feature extraction module estimates physiologically interpretable features that then get compressed by the encoding block to be next unpacked with the decoder into the delayed window of ROI sBOLD activity samples.
    \newline
    (Right) \textbf{Building blocks.} We use one layer interpretable compact block and several layers for encoder and decoder as outlined in the diagram.}
\end{figure}

BOLD from EEG Interpretable Regression Autoencoder (BEIRA) is trained to implement the following transformation: $\mathbf{X} \rightarrow \mathbf{y}$, where $\mathbf{X}$ is a $30 \times 2000$ matrix of the 30 channel EEG over the 20 s long window (sampled at 100 Hz)and $\mathbf{y}$ is the vector of the corresponding time segment of BOLD signal from one of the subcortical ROIs (sBOLD). sBOLD activity time window is shifted by a time delay which appears to be a tuned hyperparameter and lies in the 4-10 seconds range and reflects the properties of the hemodynamic response. BEIRA comprises the interpretable compact block  \cite{petrosyan2021decoding}, convolutional feature encoder and decoder. The details of the architecture are given below and visualized in Figure \ref{fig:model}.

The interpretable compact block performs spatial and frequency filtering and ensures geometric and frequency domain specificity of each of the branches. The coefficients of these filters are tuned during the training process in the context of the specific downstream task. This is done seamlessly as the entire architecture is implemented using standard DL primitives. The spatial filtering is accomplished via pointwise 1-D convolution layer followed by another depthwise 1-D convolution to perform temporal filtering followed by ReLu activation whose output is passes to another depthwise 1-D convolution for temporal smoothing. At the output of the front-end layer we obtain the timeseries reflecting the power of activity of a set of spatially localized neuronal populations. 

These timeseries are then transformed via autoencoder. The encoder comprises N encoder blocks from the \textit{wav2vec2} model \cite{wav2vec2}  followed by the dropout block \cite{dropout}. Each block allows for aggregation of time-relevant information from the continuously filtered EEG data. Each encoder block has a downsample operation which efficiently increases the corresponding receptive fields. The decoder consists of the same building block as the encoder and implements the symmetric to the encoder transformation. It upsamples output of the encoder into one ROI sBOLD signal.

\subsection{Training details}
\label{section:details}
We train and evaluate our model independently for each subject. We divide the preprocessed dataset into the training and testing sets. For the training phase we use the first 4 minutes of the preprocessed recordings. For the test we use the last 1 minute of data. We train model to reconstruct the ROI sBOLD signal by solving signal to signal translation task. We tune individual models to predict BOLD signal in each of the regions of interest. The data are parsed using 20 seconds long blocks (note also the time shift to accommodate the HRF delay). To form individual $i-th$ training sample pair $\{\mathbf{X}_i,\mathbf{y}_i\}$ we extract a 20 s long window of data formed at randomly selected staring time index $t_i$ as $\mathbf{X}_i = \mathbf{X}(:,t_i:t_i+20 *F_s)$ and $\mathbf{y}_i = \mathbf{y}(t_i:t_i+20 *F_s)$, where $F_s$ is sampling rate and equals 100 Hz.

\paragraph{Model description }

Each branch tunes to its own neuronal population using its pertinent spatial and temporal filter weights providing for geometric and frequency domain selectivity. To this end each branch performs spatial filtering operation by computing for each time point the scalar product of 30 EEG channels with the vector of 30 weights. The resultant timeseries is then processed by another branch specific convolution over the remaining temporal dimension with kernel length of 51. Then the ReLU is applied to compute the absolute values and the result is smoothed with another temporal convolution kernel of length 51. Hypothetically, the output of this front end is a set of envelopes reflecting the electrical activity of neuronal populations with geometric and dynamic properties dictated by branch specific spatial and temporal filter weights correspondingly. The spatial and temporal filters are implemented with the standard point-wise and depth-wise 1D convolutions correspondingly. The interpretable block has 16 branches and therefore may potentially serve 16 unique neuronal populations.

The encoder has 3 blocks (N=3) with the number of filters: (128, 128, 128), kernel size: (5, 5, 3), downsample: (8, 8, 4). The decoder has 3 blocks with the number of filters reduced by the factor of 4  as compared to the encoder: (32, 32, 32), kernel size: (3, 5, 5), upsample: (4, 8, 8). The dropout with 0.3 rate is applied to all autoencoder layers.

\paragraph{Loss function}

As the loss function we use a  linear combination of mean squared error(MSE) and the correlation loss:
\begin{equation}
    \label{eq:total_loss}
    \mathcal{L} = \alpha \mathcal{L}_{mse} +  \mathcal{L}_{corr},
\end{equation}
where $\alpha = 0.1 $. For correlation loss we use negative Pearson correlation coefficient.
\begin{equation}
\label{eq:corr_formula}
    \mathcal{L}_{corr} = - r(x, y) = -  \frac{{}\sum_{i=1}^{n} (x_i - \overline{x})(y_i - \overline{y})}
    {\sqrt{\sum_{i=1}^{n} (x_i - \overline{x})^2(y_i - \overline{y})^2}}
\end{equation}
Training is performed using an AdamW \cite{adamw} optimizer with a learning rate of $3 * 10^{-4}$, $ \beta_1=0.9, \beta_2=0.999$ and weight decay $3 *10^{-4}$. The batch size is  16. We use PyTorch framework to implement our models and train them using NVIDIA Tesla T4 GPU.

\section{Results}
\label{section:result}
In all our experiments described here we use Pearson's correlation coefficient as a metric for evaluation, and we run each experiment multiple times with a different seed. The correlation is then calculated between the predicted and the actual sBOLD signal. \ref{eq:corr_formula}

\subsection{Temporal relationship between EEG and sBOLD }
\begin{figure}[ht]
    \centering
    \includegraphics[width=0.6\textwidth]{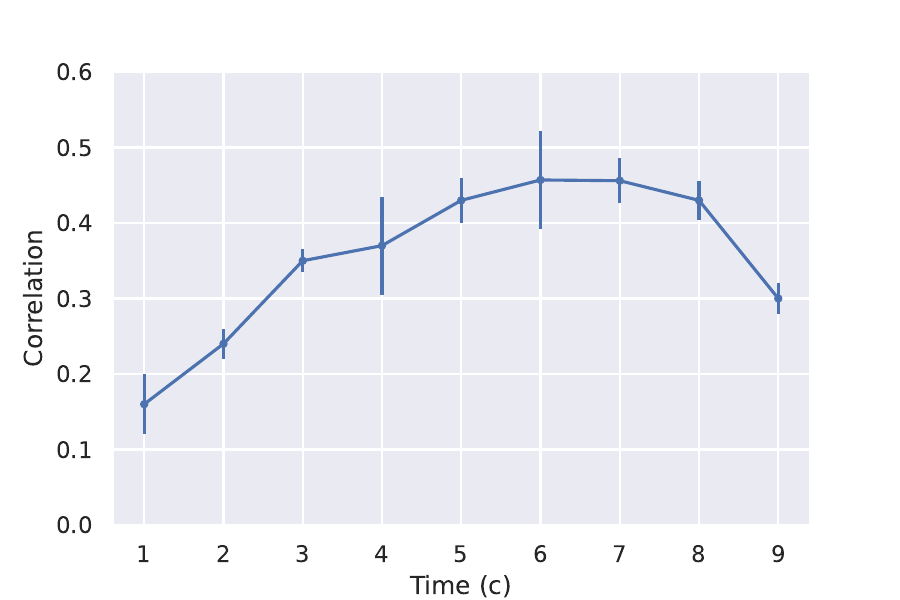}
    \caption{\label{fig:times} The dependence of the model performance on the time delay between EEG and sBOLD. The maximum performance is achieved with $t_{delay} = 6$ seconds.
     }
\end{figure}

As described earlier since BOLD signal is delayed w.r.t. EEG by several seconds and given relatively compact receptive field of our network we decided to align the two timeseries using the shift by $t_{delay}$ seconds.  In order to understand how this delay affects model performance we first evaluate our model for different time delay values. We train our model on the CWL dataset with various time delays and the average correlation is estimated across basal ganglia regions. Figure \ref{fig:times} shows an example of how the achieved correlation depends on the $t_{delay}$ hyperparameter in one subject. Estimating the optimal delay for each subject  would  ideally require more independent data and therefore we set $t_{delay}= 6$ seconds for all subsequent experiments in all other subjects. Note, that this result is  consistent with the typical shape of the hemodynamic response function (HRF). The HRF is typically asymmetric with peak latency of around 5 seconds and the relatively long late tail. Both the location of the maxima and the observed asymmetry with the faster drop near the origin appear to be consistent with the typical HRF profile.  

\subsection{sBOLD signals prediction}
In this section, we evaluate the performance of our model with a fixed time delay ( 6 seconds). For each of the 8 volunteers, we trained our model to predict six ROI of sBOLD separately. Correlations are found for different ROIs: pallidum, caudate, putamen and accumbens.

In order to compare with a baseline, we train simple regularized linear models operating on the wavelet transformed EEG data. We extract time-frequency features of EEG using the wavelet transform and flatten them into one vector. We extract 16 bands between 2 Hz and 100 Hz.  Then we train a model based on the flattened EEG features from time interval window $[t-t_{window} : t]$ seconds to predict one ROI sBOLD at moment $t$. We do not use time correction but instead use a larger window as compared to the temporal receptive field of our network. We experimented with various values for $t_{window}$ and dimension reduction schemes and regularization parameters. Finally we selected $t_{window} = 15$ second , sampling rate = 10 Hz and regularization coefficient = 0.01 which corresponded to the best performance of the linear model, see Table \ref{table:results}. The network based approach with automatically derived physiologically plausible features appeared to perform significantly better as compared to the linear model. We also observed lower variability across subjects for the BEIRA.

\begin{table}[ht]
  \caption{ \label{table:results} Comparison of different models prediction for pallidum, caudate, putamen and accumbens on the CWL dataset. Results are averaged across both bilaterally symmetric structures. We demonstrate results for one subject and  averaged across all subjects. As the model performance gauge we use Pearson's correlation coefficient between the EEG-derived and the actual ROI sBOLD timeseries. 
  }
  \label{sample-table}
  \centering
  \begin{tabular}{llllll}
    \toprule
    \multicolumn{4}{r}{CWL dataset}\\                  
    \cmidrule(r){2-5}
    Method & Pallidum & Caudate & Putamen & Accumbens & Average \\
    \midrule
    \midrule
    \textbf{All subjects} & & & & & \\
    Ridge linear model     & 0.05 $\pm$ 0.15 & 0.34 $\pm$ 0.14 & 0.38 $\pm$ 0.24 & 0.26 $\pm$ 0.10 &  0.24 $\pm$ 0.11 \\
    \midrule
    \textbf{BEIRA} & 0.37 $\pm$ 0.04 & 0.47 $\pm$ 0.14 & 0.48 $\pm$ 0.16 & 0.42 $\pm$ 0.05 &\textbf{0.44 $\pm$ 0.05}\\
    \midrule
    \midrule
    \textbf{One subject} & & & & & \\
    Ridge linear model     &0.01  & 0.40 & 0.48 & 0.10& 0.25$\pm$ 0.20\\
    \midrule
    \textbf{BEIRA}    & 0.39 $\pm$ 0.01  & 0.62 $\pm$ 0.05& 0.55 $\pm$ 0.04 &0.37 $\pm$ 0.01 & \textbf{0.48 $\pm$ 0.02} \\
    \bottomrule
  \end{tabular}
\end{table}

\begin{figure}[ht]
    \centering
    \includegraphics[width=0.7\textwidth]{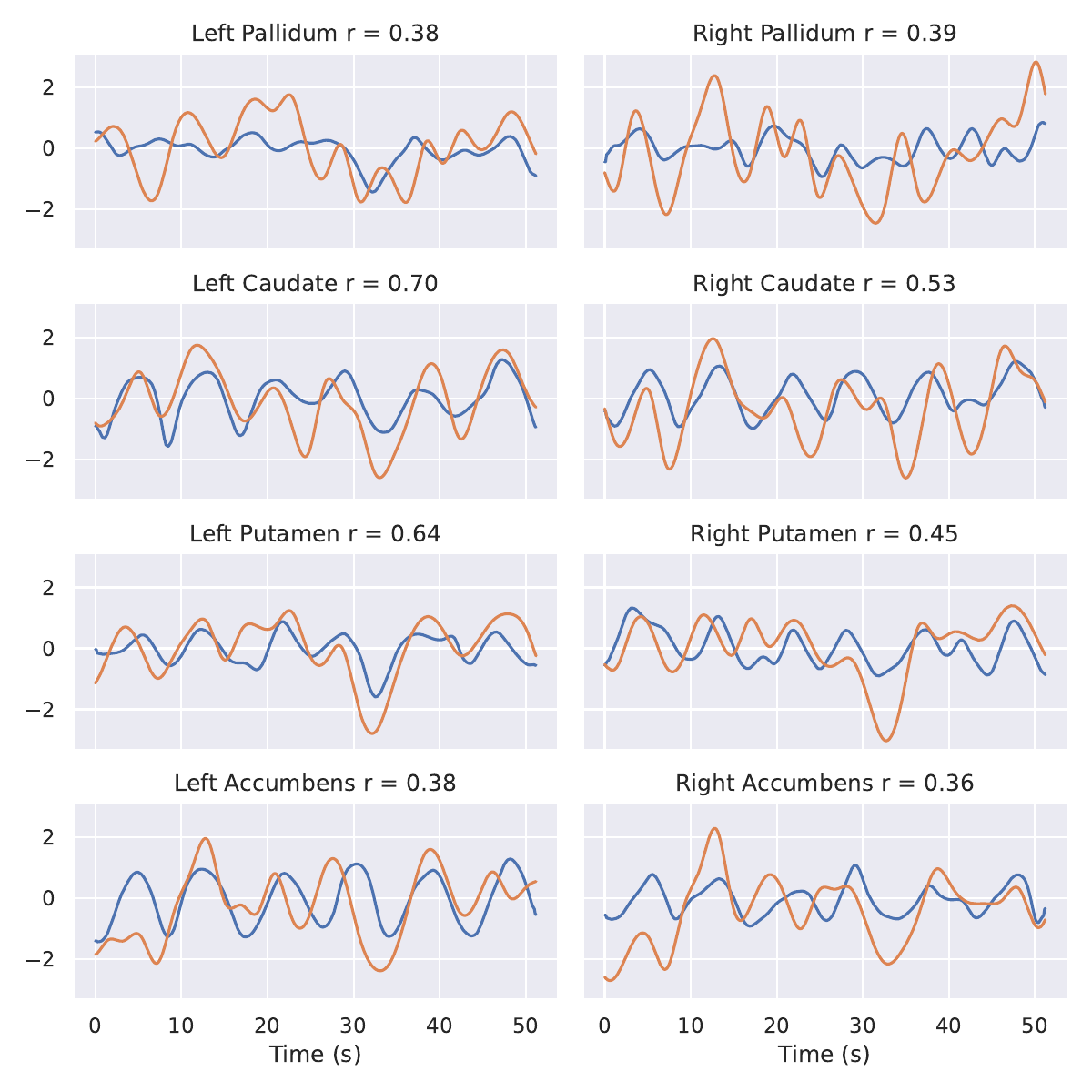}
    \caption{\label{fig:ts_plot} Real and predicted sBOLD time series of different RoI for one patient from CWL dataset: pallidum, caudate, putamen, accumbens. Yellow is real, blue is prediction. x axis - Time seconds. y axis - sBOLD activity.
     }
\end{figure}

Next in Figure \ref{fig:ts_plot} we show the predicted by BEIRA and actual ROI sBOLD timeseries for eight subcortical ROI regions in one subject. 

We observe a reasonable agreement between the actual and predicted ROI sBOLD traces which significantly exceeds chance level. The chance level accuracy can be partly appreciated by exploring the dependence of the average accuracy on the $t_{delay}$ hyperparameter shown in Figure \ref{fig:times}. This plot illustrates that the observed performance appears to be naturally dependent on the mutual arrangement of the EEG and sBOLD timeseries and the best accuracy corresponds to the typical HRF's peak latency. It can also be seen that when the time shift leaves physiologically plausible range the performance drops from 0.5 to around 0.15.        

We note that we operated using a very small amount of data, only 4 minutes of data was used for training our model. This was nevertheless sufficient to demonstrate the basic feasibility of decoding resting state basal ganglia BOLD signal from the surface EEG recordings. The EEG data we used were recorded in a continuous fMRI + EEG paradigm and therefore appear to be quite seriously contaminated with gradient artifacts. The removal of these artifacts is known to significantly corrupt the EEG data especially their higher frequency components. In the future the model can be trained using the intermittent fMRI + EEG paradigm, where the EEG  used to decode the BOLD signal is recorded during the time intervals when the gradient coils are turned off. This can be easily accomplished given the natural delay of several seconds between the EEG and BOLD, see Figure \ref{fig:times}.


\subsection{Model interpretation}
\begin{figure}[ht]
    \centering
    \includegraphics[width=0.6\textwidth]{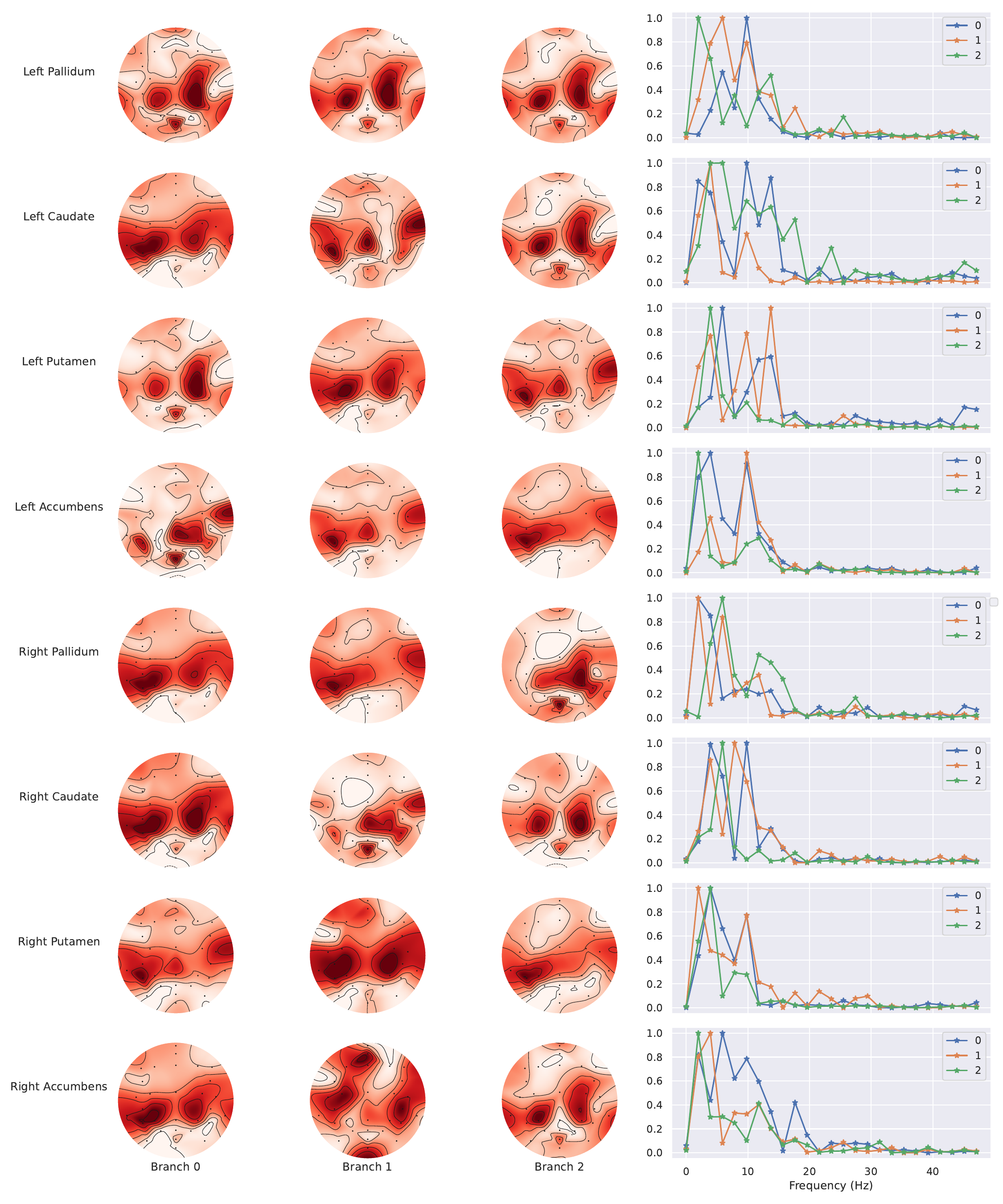}
    \caption{\label{fig:topomap} \textbf{EEG topomaps and frequency patterns} We provide spatial filters interpretation and extract spatial patterns which demonstrate influence of different electrodes in prediction. Intensity of the color means contribution of area. Additionally, we show frequency patterns for each ROI sBOLD.
     }
\end{figure}

Next, we evaluate spatial and frequency patterns of the trained models (Figure \ref{fig:topomap}). We use the method described in \cite{petrosyan2021decoding} for compact interpretable block interpretation. This method allows us to extract spatial and frequency patterns for each branch of our front-end network.  

To find out which branches of compact block have the greatest contribution to the prediction of the model, we have applied the layer$\times$gradient method which is based on Saliency map principle \cite{saliency}. We element-wise multiply the layer's activation with gradients of the output with respect to the given layer. Feature importance is then determined by the absolute value of these coefficients. Then feature importance is used for choosing the first few most influential branches for  interpretation. In Figure \ref{fig:topomap} we demonstrate spatial and frequency patterns for the top 3 most important branches.

We observe for some ROIs spatial patterns appear similar but the frequency domain portraits differ and cover different sub-bands. This configuration may support calculation (performed by the subsequent layers) of the extent to which the spectrum of activity of a specific neuronal population extends into higher frequencies which is known to be one of the EEG correlates of the fMRI signal \cite{hemoeeg}. Also, the topographies mainly emphasize motor cortex which is consistent with the functional role of basal ganglia and the presence of the corresponding interconnections leading to motor cortical areas \cite{purves2001neuroscience}.

\section{Summary and Discussion}

\label{summary}

The proposed model predicts the sBOLD signal of multiple subcortical structures based on the multichannel EEG data. To our knowledge, this is the first study in which the following subcortical areas were successfully predicted: the pallidum, caudate, putamen and accumbens using an interpretable deep neural network which did not require explicit feature engineering. On the contrary, after training the network we interpreted the obtained decision rule which allowed us to make statements regarding the intricate temporal, frequency and spatial relationships of sBOLD and EEG data. The observed relationships appear to support the existing theoretical models linking EEG and BOLD.  With this work we not only contribute towards creation of wearable and low cost  subcortical activity scanners but also showcase an automatic knowledge discovery process facilitated by deep learning technology in combination with an interpretable domain constrained architecture and the appropriate downstream task. 
 
The proposed approach after it is further developed and properly tested will allow us to discover the exact details of EEG with subcortical BOLD. Ultimately, it will lead to a set of algorithms and rules to predict basal ganglia BOLD activity from the low cost and wearable scalp recordings. We foresee how this technology is used for monitoring basal ganglia activity in patients at remote sites after the subject-specific model has been created. 

\section{Acknowledgment}
This work is supported by the Center for Bioelectric Interfaces NRU HSE, RF  Government grant, AG. No. 075-15-2021-624

\bibliographystyle{plain} 
\bibliography{refs} 



\appendix

\end{document}